# Antimony thin films demonstrate programmable optical non-linearity

*Zengguang Cheng[1,2*], Tara Milne[1], Patrick Salter[3], Judy S. Kim[1], Samuel Humphrey[1], Martin Booth[3], and Harish Bhaskaran[1*]*

[1] Department of Materials, University of Oxford, Parks Road, Oxford OX1 3PH, UK

[2] State Key Laboratory of ASIC and System, School of Microelectronics, Fudan University, Shanghai 200433, China

[3] Department of Engineering Science, University of Oxford, Parks Road, Oxford OX1 3PJ, UK

[*] Corresponding authors

E-mail: zgcheng@fudan.edu.cn (Z.C.)

E-mail: harish.bhaskaran@materials.ox.ac.uk (H.B.)

**The use of metals of nanometer dimensions to enhance and manipulate light-matter interactions for a range of emerging plasmonics-enabled nanophotonic and optoelectronic applications is an interesting, yet not highly explored area of research outside of plasmonics[1,2]. Even more importantly, the concept of an active metal, i.e. a metal that can undergo an optical non-volatile transition has not been explored. Nanostructure-based applications would have unprecedented impact on both the existing and future of optics with the development of active and nonlinear optical tunabilities in single elemental metals[3-5]. Compared to alloys, pure metals have the material simplicity and uniformity; however single elemental metals have not been viewed as tunable optical materials, although they have been explored as viable electrically switchable materials. In this paper we demonstrate for the first time that antimony (Sb), a pure metal, is optically distinguishable between two**

**programmable states as nanoscale thin films. We then show that these states are stable at room temperature, and the states correspond to the crystalline and amorphous phases of the metal. Crucially from an application standpoint, we demonstrate both its optoelectronic modulation capabilities as well as speed of switching using single sub-picosecond (ps) pulses. The simplicity of depositing a single metal portends its potential for use in applications ranging from high speed active metamaterials to photonic neuromorphic computing, and opens up the possibility for its use in any optoelectronic application where metallic conductors with an actively tunable state is important.**

**Introduction**

A large array of applications ranging from optical coatings[6,7], plasmonic antenna[8,9], metasurfaces[10,11], high resolution imaging[12], biosensors[13] to integrated photodetectors and modulators[14,15] would benefit from active tunable optical properties of metallic thin films and nanostructures. Existing technologies to achieve active tunability either by incorporating tunable electro-optical materials[16], laser post-processing[17] or electrolyte gating[5], are limited to low speed, irreversible or low energy efficiency. Although the phase transition of metals between amorphous and crystalline structure have been studied since 1960s[18-20], there have been no reports of single elemental metals having an actively tunable optical property. Indeed, even electronically, it was only recently amorphous states of single element metals have been obtained by nanosecond (ns) electrical pulse melt-quenching[21-23].

In this paper, we report on antimony (Sb), which when configured as a thin film of nanometric dimensions, behaves reliably as a tunable optical material. Such a functionality allows us to explore it use in a range of optical and optoelectronics applications as we demonstrate. The use of optical property contrast between two phases is not unknown in the context of a class of alloys known as phase-change materials; it is no accident that those very properties of those materials have seen exploitation in photonic applications, including reflective nano-displays[24], tunable emitters and absorbers[25,26], reconfigurable meta-photonics[27-29], and integrated phase-change photonics[30-35], accompanied by the development of specialized optical PCMs[36-38] and nanostructured optoelectronic devices[39,40]. However, a common limitation with alloys is miniaturization, where maintaining compositional integrity is difficult at reduced dimensions.

We show that monoatomic metal materials with tunable non-volatile optical properties could benefit photonic applications based on active metallic nanostructures and miniaturized metallic memories. Crystalline Sb is a single element metal and its amorphous phase has been obtained by careful deposition of thin film[20,41] or electrical pulse switching[22,23] with significantly decreased electrical conductivity working as a semiconductor. During the metal-insulator transition of Sb[42], a substantial change of the free carrier absorption will result in a significant contrast in its optical loss. Therefore, an optical property change of pure Sb could be expected during the phase transition, yet this has never been studied.

Here, we systematically studied the phase transition of ultra-thin pure Sb in the optical domain using optical, electrical and structural characterizations. We demonstrate that pure Sb is a promising tunable optical material with significant non-volatile change in optical properties, especially the extinction ratio ($k$), between the amorphous and crystalline phases. We further demonstrate that pure Sb can be amorphized by a single shot femtosecond (fs) pulsed laser with a tunable retention time of the switched amorphous phase. As we further show, this has significant applications in reflective displays and potentially in future integrated photonics.

**Material characterizations of ultra-thin Sb films**

First, we investigate the dependence of Sb thicknesses ($t_{Sb}$) on optical constants, refractive index ($n$) and extinction ratio ($k$), as $n$ and $k$ are the key parameters for optical applications. Sb films with no capping layers were directly sputtered on silicon wafers and then characterized by ellipsometry measurements from which optical constants are determined (Fig. 1, a and b). For thin film Sb ($t_{Sb} \leq 11$ nm) as deposited, the dependence of refractive index $n_a$ is weak in the ultraviolet and visible regimes (200~800 nm) with an increasing of $n_a$ vs. $t_{Sb}$ in the infrared (Fig. 1a). Similarly, the extinction ratio $k_a$ increases monotonically with $t_{Sb}$ from visible to infrared yet on a larger scale (Fig. 1b). After annealing on a hotplate at 270 °C for 10 mins, the same samples were further investigated by ellipsometry with optical constants shown in Fig 1, c and d. Optical constants of ultra-thin c-Sb (3 and 4nm) do not follow the trend of thicker samples, with significantly smaller extinction ratio $k_c$ values (Fig. 1d and Supplementary Fig. 1). Furthermore, when compared with a-Sb samples, the refractive index change $|\Delta n|$ ($|n_c - n_a|$) after crystallization is less than 1.5 (Fig. 1e), whereas the change in its extinction ratio $|\Delta k|$ ($|k_c - k_a|$, Fig. 1f) is considerable with a maximum value over 3 in telecom wavelength bands (1.5-1.6 μm). The $|\Delta k|$ of Sb is much larger than that for GST and other PCMs (which is between 0.15 and 1.8 at 1.55 μm)[43]. With increasing $t_{Sb}$ (up to 20 nm) optical constants approach those of bulk Sb with less changes after annealing (Supplementary Fig. 1), which demonstrates that optical contrast of the phase transition can only be obtained from a thickness-confined thin film, less than 15 nm for the specific structure studied here. In spite of the fact that the thickness-dependent optical property has been demonstrated in 2D materials due to the quantum confinement and the interlayer coupling[44,45], this has not been widely reported in thin film Sb.

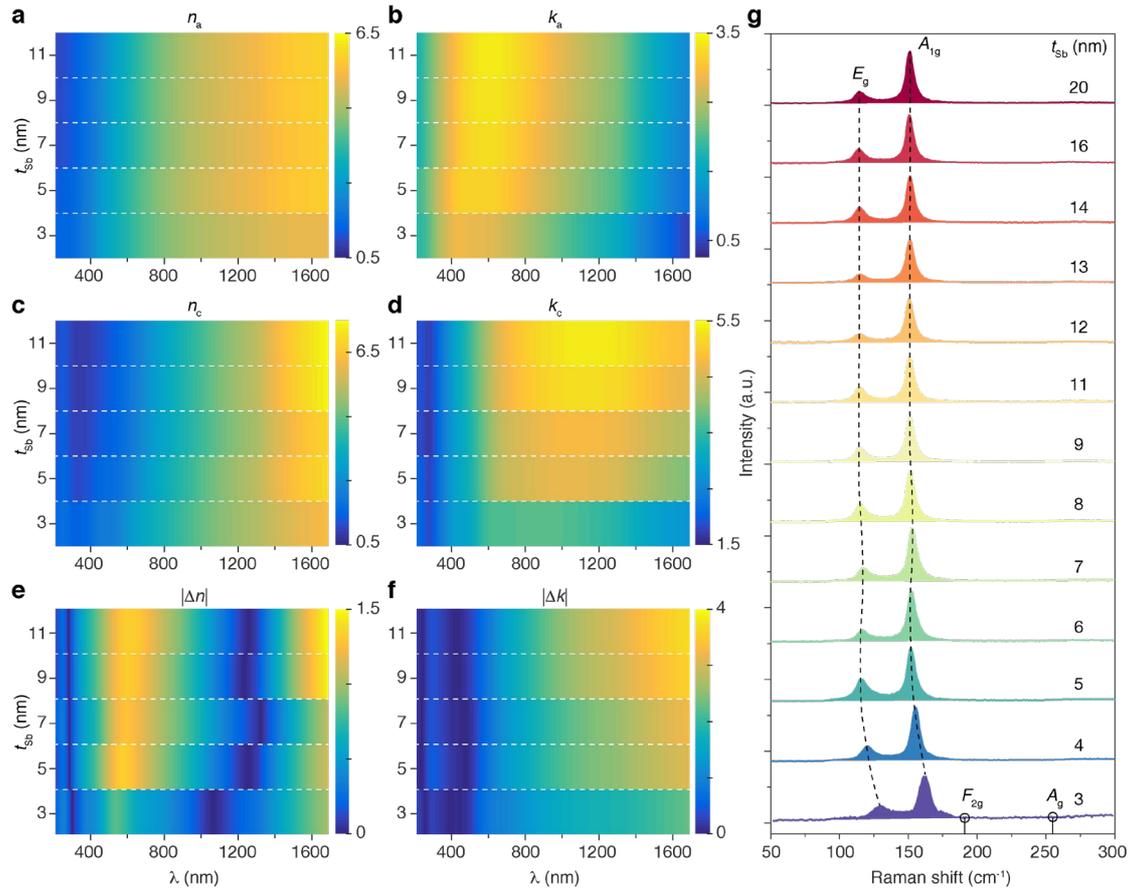

**Figure 1 | Thickness dependent of optical and structural properties of Sb.**
(**a** and **b**) The spectra (from ultraviolet to near infrared) of the refractive index $n_a$ (a) and extinction ratio $k_a$ (b) of thin film Sb with different thicknesses ($t_{Sb}$) as deposited on silicon wafers measured by spectroscopic ellipsometry. (**c** and **d**) The spectra of $n_c$ (c) and $k_c$ (d) of the same Sb samples in (a and b) after annealing on a hotplate (270 °C for 10 mins). (**e** and **f**) The absolute change of refractive index $|\Delta n|$ ($|n_c - n_a|$) (e) and extinction ratio $|\Delta k|$ ($|k_c - k_a|$) (f) of Sb upon annealing, calculated from (a and c) and (b and d) respectively. (**g**) Raman spectra of annealed Sb films with different $t_{Sb}$. $E_g$ and $A_{1g}$ vibration modes are denoted by the dashed lines. Typical vibration modes $F_{2g}$ and $A_g$ of $Sb_2O_3$ are illustrated by the circles. The Raman spectrum intensity of 3 nm Sb (purple) has been enlarged by two times for clarification.

Additionally, Raman spectra of c-Sb samples have been investigated in Fig. 1g. Typical in-plane ($E_g$) and out-of-plane ($A_{1g}$) vibrational modes of Sb are denoted in the figure. For $t_{Sb}$ larger than 9 nm, Raman peaks for $E_g$ and $A_{1g}$ are at ~114 cm$^{-1}$ and ~151 cm$^{-1}$, consistent with that in bulk Sb[46]. When $t_{Sb}$ gradually decreases to 3 nm, both peaks for $E_g$ and $A_{1g}$ blue shifted to larger wavenumbers. Similar phenomena have been reported in 2D antimonene[46,47] relevant with local lattice contractions. It is worth noting that no antimony oxide ($Sb_2O_3$) Raman peaks at ~191 cm$^{-1}$ and ~255 cm$^{-1}$ were observed in our samples which confirms optical properties change of Sb upon annealing is due to the phase change rather than from any oxidation. On the other hand, the Raman spectrum of Sb before annealing is insensitive to the thickness and clearly shows an amorphous phase for all samples (Supplementary Fig. 2), indicating that all Sb films undergo the phase transition from amorphous to crystalline upon annealing. However, only thin film Sb ($t_{Sb}$ < 15 nm) has a significant change in optical and electrical properties. It has been suggested that local clusters of Sb ($Sb_1$ or $Sb_4$) are important to the electrical properties of ultra-thin films and Raman spectra but has little effect to the electrical properties of thick Sb films[20]. The amorphous phase of thick Sb (> 15 nm) behaves more like a metallic glass[21] rather than a semiconducting material.

To identify the crystal structure of Sb before and after the thermal annealing, transmission electron microscopy (TEM) and selected area electron diffraction (SAED) were implemented. Sb was sputtered as a 5 nm film on carbon films supported by copper grids. TEM of as-deposited Sb morphology is elucidated in Fig. 2a showing an amorphous disordered structure, which is confirmed by the diffusion halo pattern of SAED (Fig. 2b). After thermal annealing (270 °C, 10 mins), the morphology of Sb changed (Fig. 2c) with clear spot patterns in SAED (Fig. 2d) verifying its hexagonal crystalline structure. Two groups of diffraction patterns, corresponding to zone axis of [0 0 1] and [-1 0 0], were observed in a 200 nm diameter selected area region. The TEM image in Fig. 2e shows crystalline planes perpendicular to the [1 -2 0] direction. The interplanar spacing was measured as 2.2 Å using the Fourier Transform (FT) spots (inset of Fig 2e). From the SAED pattern in Fig. 2d, we further calculated the crystalline constants $a$ and $c$ of the rhombohedral crystalline structure of Sb (Fig 2f) as: $a$ = 4.55 Å and $c$ = 11.53 Å ($c/a$ = 2.53), consistent with bulk Sb ($a$ = 4.31 Å, $c$ = 11.27 Å, $c/a$ = 2.61)[48]. Although the thickness of Sb layer is only 5 nm here, its structure (orientation) is different from few layer 2D Van der Waals Sb (antimonene). The diffraction spots (0 0 6) and (0 0 -6) corresponding to $c$-planes are clearly shown in Fig. 2d which are typically missing in antimonene whose $c$-planes are parallel to substrates and perpendicular to the electron beam[46,47].

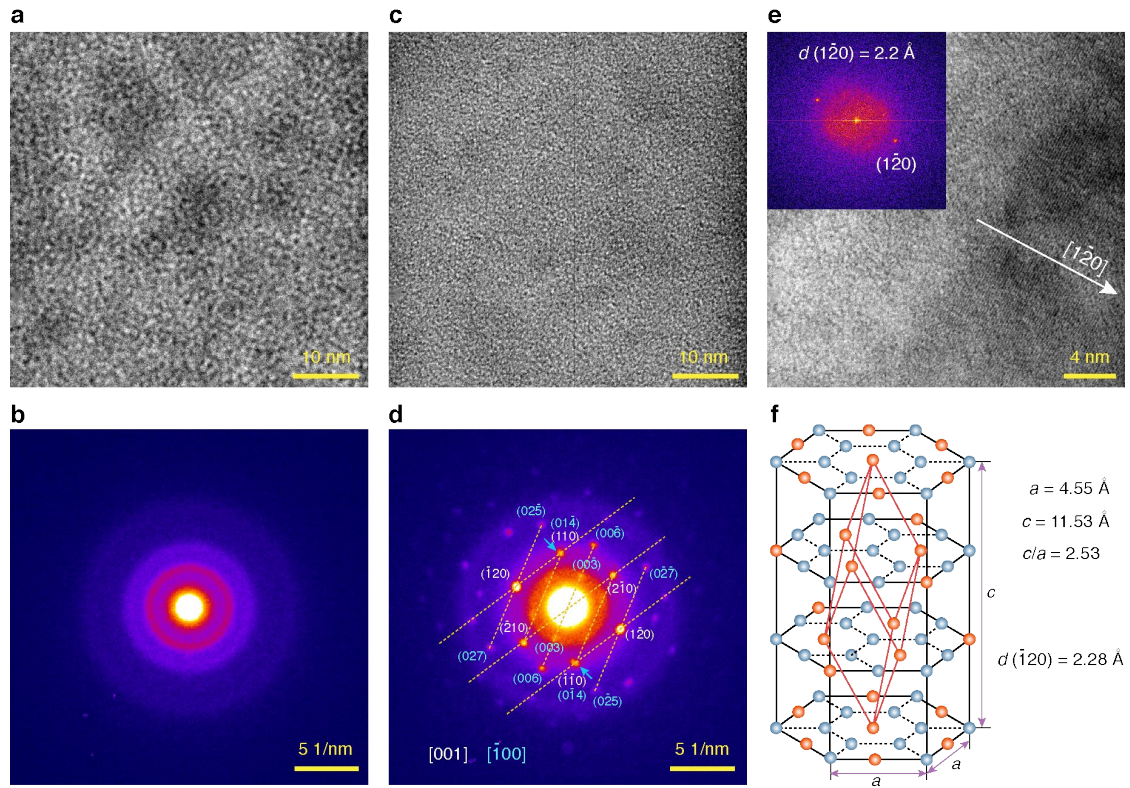

**Figure 2 | Transmission Electron Microscopy characterization of Sb.**
(**a**) TEM image of 5 nm thick Sb layer as deposited has an amorphous structure. (**b**) Selected area (344 nm diameter) electron diffraction (SAED) of the as-deposited Sb, corresponding to the sample in (a). (**c**) TEM image of the same Sb sample in (a) after thermal annealing. (**d**) SAED (200 nm diameter region) of the annealed Sb sample in (b) has formed diffraction spots from crystalline planes. Two patterns, corresponding to the zone axis of [0 0 1] (white) and [-1 0 0] (cyan), are overlapped. Miller indices of crystal planes with high symmetry are labeled. The arrows show merged diffraction spots coming from different planes. (**e**) TEM image of the annealed Sb with visible crystalline planes along the [1 -2 0] direction. **Inset**: Fourier transform (FT) of the whole image showing reflections (1 -2 0) with an interplane spacing (*d*) of 2.2 Å. (**f**) The hexagonal unit cell of the rhombohedral crystalline structure of Sb. The red lines show the primitive rhombohedral unit cell. *a* and *c* are the measured crystal constants in the hexagonal unit cell.

**Applications in strongly interfering optics**

Next, we explore how thin film Sb responds within strongly reflecting thin film structures. To do this, we demonstrate a reflective display structure[24] incorporating thin film Sb PCMs. As shown in Fig. 3a, a thin film Sb is sandwiched between two indium tin oxide (ITO) layers which have been deposited in sequence on a platinum (Pt) mirror. The thickness of Sb is fixed at 5 nm with a 15 nm top ITO capping layer. The reflective color of the sample is highly dependent on the thickness ($t_{ITO}$) of the bottom ITO. We fabricated reflective display samples on silicon wafers with varying thicknesses of the bottom ITO: $t_{ITO}$ = 50, 75, 100, 125 and 150 nm (Fig 3b). The as-deposited thin Sb layer confined by ITO is in the amorphous phase (a-Sb). Reflective display samples incorporation a-Sb layers in the top panel of Fig. 3B show the reflective color changes from dark blue to bright yellow with increasing $t_{ITO}$. In order to achieve fully crystalline Sb (c-Sb), we thermally annealed samples at 270 °C for 5~10 mins on a hot plate to produce reflective colors shown in the bottom panel of Fig. 3b. A significant color change was observed after the thermal annealing in all samples except for $t_{ITO}$ = 150 nm. Both samples including a-Sb and c-Sb layers have been kept in atmosphere for 6 months without any color degradation. The color changes of these samples were further validated by the measured reflection spectra (Fig. 3c), where the maximum reflection peak is seen to shift red with the increasing of $t_{ITO}$ accompanied by a notable discrepancy of the spectra before and after annealing. This observation is consistent with the simulated reflective spectra in Fig. 3d, using a transfer matrix computational method[15] with optical constants for 5 nm Sb obtained by ellipsometry measurements (Supplementary Fig. 1). In addition, similar Sb stacks can be employed in reconfigurable metasurfaces[28] holographic displays[49] by optimizing the structure of the stack, with substantial applications in spatial light modulators and head-up displays for virtual and augmented reality.

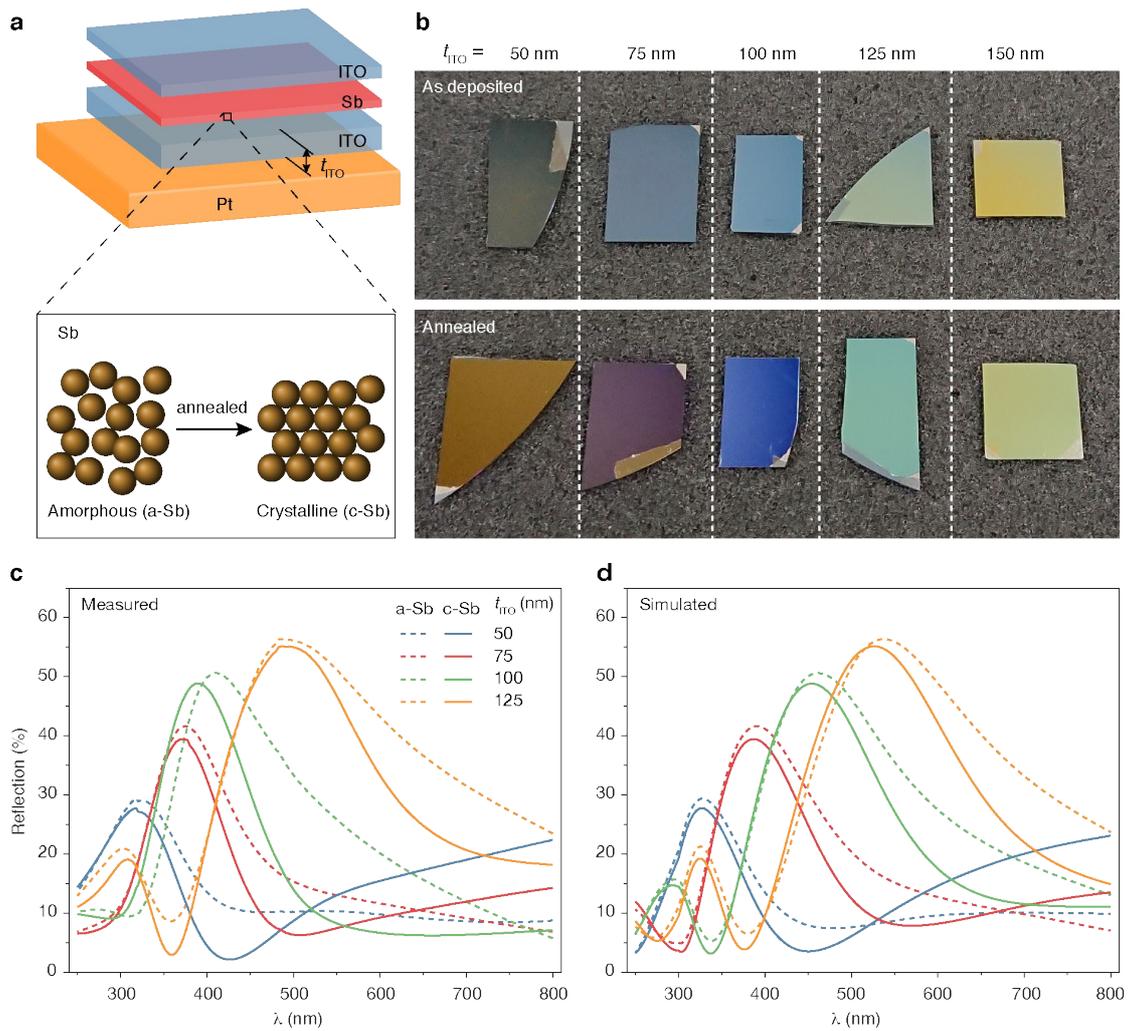

**Figure 3 | Switchable reflective stacks using ultra-thin-film Sb.**
(**a**) Structure of reflective display based on phase-change Sb sandwiched between two ITO layers (ITO/Sb/ITO) on top of a Pt mirror. The phase-change Sb layer can be switched from amorphous to crystalline through thermal annealing. (**b**) Optical images show typical display samples with different thicknesses ($t_{ITO}$) of the bottom ITO layer, while Sb and the top ITO are fixed at 5 nm and 15 nm respectively. The samples have amorphous (top row) and crystalline (bottom row) Sb layer. (**c** and **d**) Measured (c) and simulated (d) reflection spectra of samples corresponding to (b). Each measured spectrum curve was normalized to the peak of the corresponding simulated curve.

**Optoelectronic modulation of Sb**

We then explore how thin film Sb responds in the optoelectronic domain. To study this, we carry out electrical switching of the materials at nanoscale to investigate whether this results in optical contrast. Conductive atomic force microscopy (CAFM) is a versatile method used for nanoscale crystallization[24,50]; we employ CAFM to crystallize a-Sb with similar thin film structures in Fig. 1. As shown in Fig. 4a, the Sb layer is encapsulated by the top and bottom ITO layers working as electrical contacts for Sb. The bottom ITO layer above the Pt layer is grounded through a protective resistor ($R_S$). The conductive AFM tip is in contact with the top ITO layer with DC voltages applied, resembling the vertical structure of standard phase-change memory cell. The current $I_S$ passing vertically through Sb is monitored while the biased voltage $V_B$ is varied. The current $I_S$ is negligible at small $V_B$ and rapidly increases to a high conductive state when $V_B$ reaches a threshold voltage ($V_{th}$) indicating a localized crystallization. We implemented the measurement over 20 different positions on the sample; this is shown in Fig. 4b, where the conductivity change is over 2 orders of magnitude during the switching consistent with previous studies on electrical switching of Sb[20,22,23], although with a wide distribution of $V_{th}$ (inset of Fig. 4b). Next, grey scale images were patterned on Sb stacks by modulating $V_B$ on AFM tip while raster scanning the samples. For the stack of 15 nm ITO/5 nm Sb/100 nm ITO/Pt, the pixel color was switched from pale blue (a-Sb) to dark blue (c-Sb) by CAFM with the optical image taken in Fig. 4f, corresponding to the original picture in Fig. 4c. By reducing the thickness of Sb and further optimization of the stack (15 nm ITO/3 nm Sb/50 nm ITO/Pt), we have reached significant improvement of the contrast of the switched images as shown in Fig. 4, g and h, from original pictures in Fig. 4, d and e respectively. With this design, grey scale images have been perfectly replicated on the stack with a high resolution (< 200 nm/pixel). The preservation of image detail is also very good (Fig. 4, i-l). Importantly, colors between a-Sb and c-Sb inferring intermediate phases have been achieved due to different bias voltages.

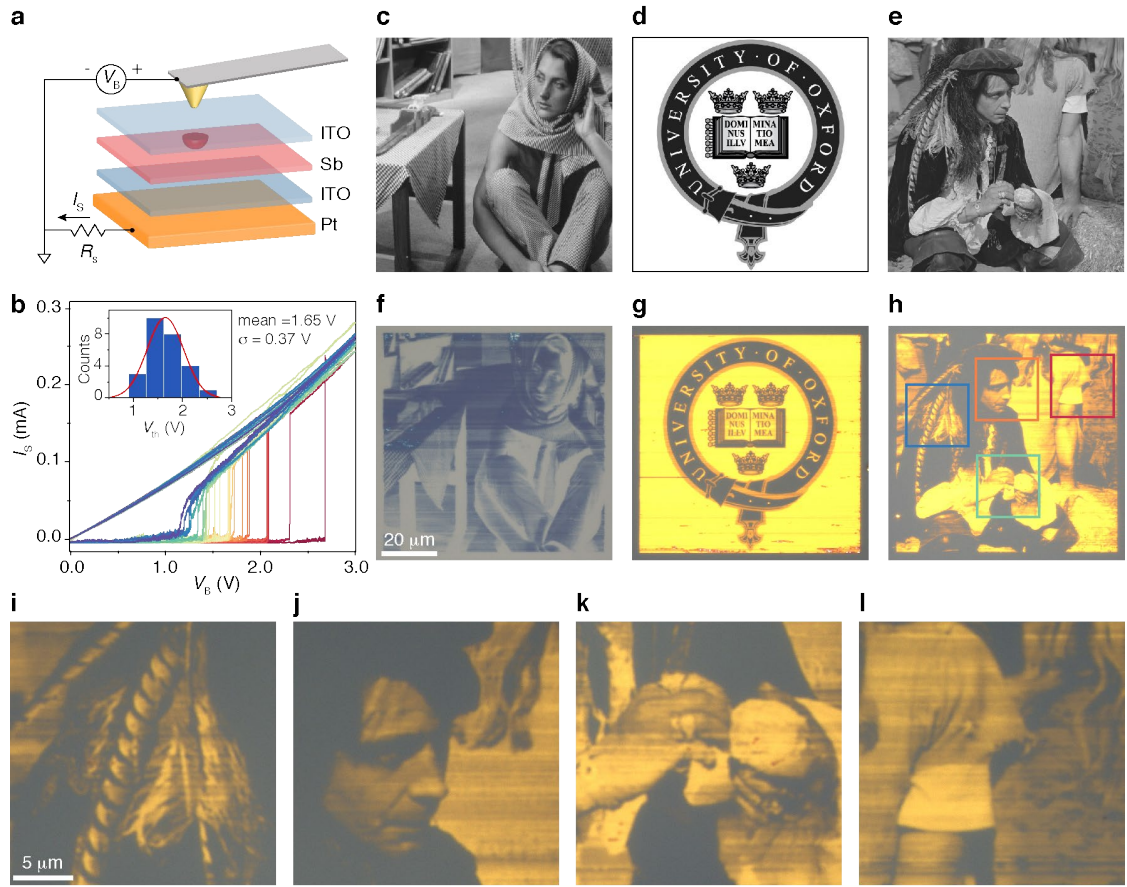

**Figure 4 | Electro-optical switching of Sb using Conducting AFM.**
(**a**) Schematic of the electrical switching of Sb using CAFM. The Sb film is sandwiched between two ITO layers above a Pt mirror. The conductive probe of the CAFM is biased using a DC voltage ($V_B$) while contacting or scanning the Sb sample. The Pt substrate is grounded via a resistor $R_S$ (3 kΩ) to limit the current ($I_S$) passing through the probe and the sample. (**b**) Static measurement of $I_S$ while sweeping $V_B$ on the probe that is in contact with different locations of the sample. **Inset**: Histogram distribution of the threshold voltage $V_{th}$ of the switching during voltage sweeping. (**c-e**) Original grey scale (8-bit, 256 × 256) images used to modulate $V_B$. (**f**) The optical image of Sb sample switched by CAFM, corresponding to the image in (c). The Sb sample structure is 15 nm ITO/5 nm Sb/100 nm ITO/ Pt (from top to bottom layer). (**g** and **h**), The optical image of Sb sample switched by CAFM, corresponding to the images in (d and e) respectively. The Sb sample structure is 15 nm ITO/3 nm Sb/50 nm ITO/Pt. (**i-l**), Optical images show zoomed-in regions of the switched areas: blue (i), orange (j), green (k) and red (l) boxes in (h).

**Fast and reversible switching of Sb using fs laser**

Finally, we turn to studying both the reversibility of switching in these materials and their dynamic speed. Particularly for emerging applications in photonic computing, sub-nanosecond (ns) switching speeds are required, and faster speeds approaching picoseconds are highly desirable, which most PCMs are unable to reach. Earlier work demonstrated that the amorphization of c-Sb occurs using nanosecond electrical pulses at cryogenic temperature; this indicates that a much faster process is necessary for the amorphization at room temperature[22]. For this reason, we chose a femtosecond pulsed laser, to optically switch Sb. The optical switching setup is illustrated in Fig. 5a, a regeneratively amplified Ti: Sapphire femtosecond laser ($\lambda$ = 790 nm, 1 kHz repetition rate, pulse 200 fs) was focused on Sb samples through a 10 × objective lens. Sb samples were mounted on a positioning stage for raster switching a large area. Similar to the electrical switching in Fig. 4, our sample is based on the ITO/Sb/ITO/Pt stack structure providing a good reflective color contrast that is readily observed using optical microscopes. A Sb stack sample (15 nm ITO/3 nm Sb/50 nm ITO/Pt) has been completely crystallized by thermal annealing. As shown in Fig. 5b, a single fs laser pulse was used to amorphize the c-Sb stack sample with various pulse energies ($E_p$). The switched region has a circular shape with a color change (to dark blue). The size of the switched region gradually increases with $E_p$, until at very high power we ablate the entire stack (eventually exposing the underlying Pt at high $E_p$). The switched regions were further characterized as a-Sb by Raman spectra (Supplementary Fig. 3 and 4), confirming that the color changes are a consequence of amorphization. Subsequently, large areas of a-Sb have been switched via the scanning of the sample stage while using single fs pulse with a moderate energy ($E_p$ = 0.56 nJ). Two typical amorphized regions (a-Sb1 and a-Sb2) are shown in Fig. 5c, with local reflection spectra measured in Fig. 5e suggesting robust and reproducible amorphization.

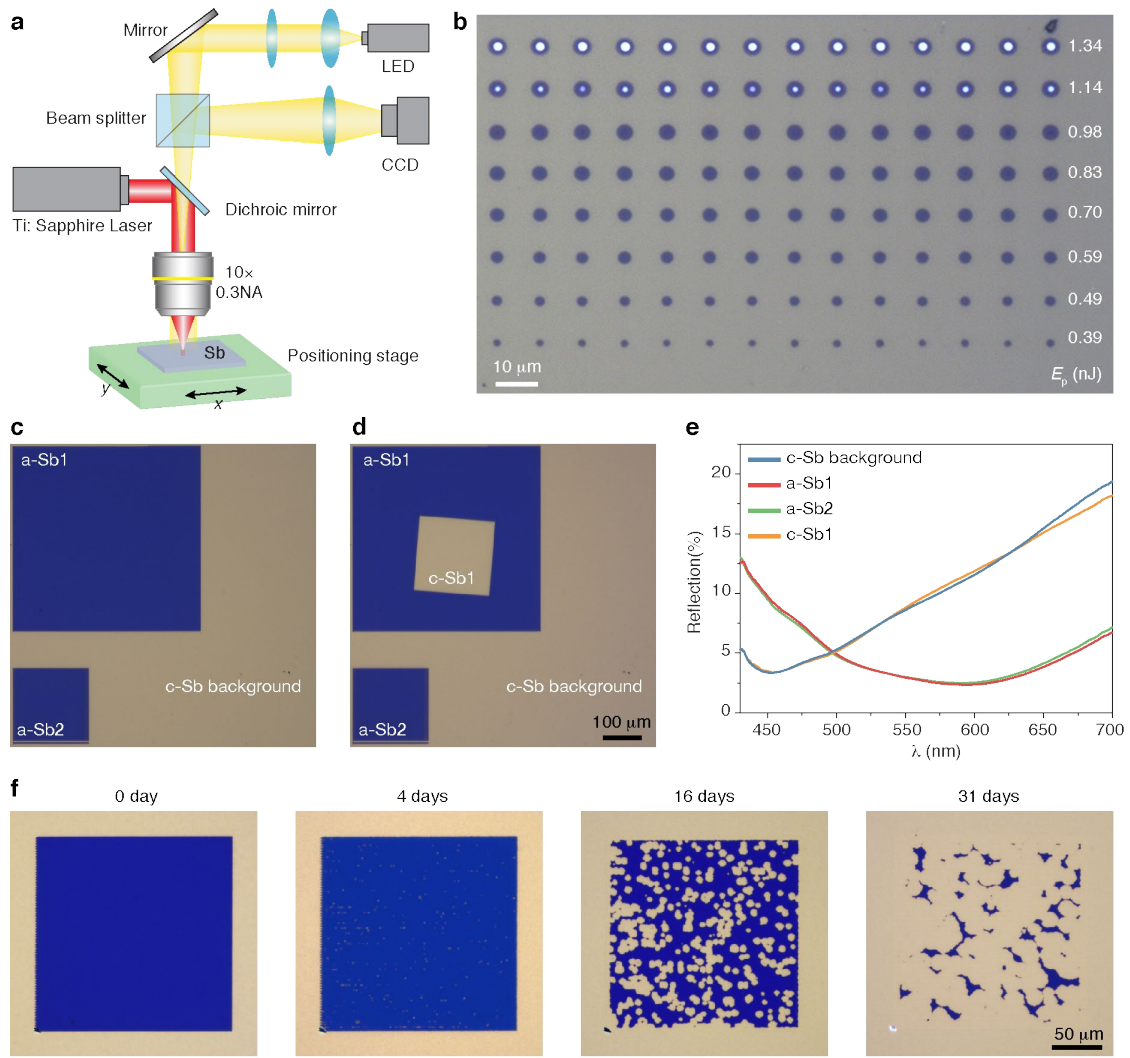

**Figure 5 | Ultrafast optical switching of Sb.**
(**a**) Schematic of optical switching of Sb using fs laser. (**b**) Optical image shows amorphized regions (blue disks) using single laser pulse (200 fs) with increasing energy $E_p$ (from bottom to top). (**c**) Large area switching of crystalline Sb sample (c-Sb background) through single fs pulse (200 fs, $E_p$ = 0.56 nJ) while raster scanning the sample (moving speed 500 μm/s). a-Sb1 and a-Sb2 are switched areas with different sizes. (**d**) Recrystallization (c-Sb1) of the amorphized region a-Sb1 in (c) with multiple fs pulses (200 fs, $E_p$ = 29 pJ, 80 MHz) while translating the sample at 200 μm/s. (**e**) Reflection spectra of different locations of the sample in (d). (**f**) The stability of the amorphized region by fs laser switching. Optical images show the amorphized region after different aging times in ambient conditions at room temperature. The Sb sample used is 15 nm ITO/3 nm Sb/50 nm ITO/Pt.

To demonstrate reversible switching, i.e. the recrystallization of switched a-Sb regions, a pulse train from the Ti: Sapphire oscillator with 80 MHz repetition rate and significantly lower pulse energy ($E_p$ = 29 pJ) was used as the laser source with the recrystallized region (c-Sb1) shown in Fig. 5d. The reflection spectra (Fig. 5e) and Raman spectra (Supplementary Fig. 5) of the recrystallized c-Sb1 are consistent with the background c-Sb. Since amorphized Sb has a strong tendency to recrystallize, we monitored the evolution of an optically switched a-Sb region at room temperature (RT, ~24 °C), as shown in Fig. 5f and Supplementary Fig. 6. The switched a-Sb region was very stable over 36 hours; initial nucleation of the c-Sb after 4 days was observed then followed by a gradual growth of nucleated regions. The whole recrystallization process for the amorphous region took more than one month. By slightly increasing the thickness of Sb to 5 nm, the initial nucleation in the optically switched a-Sb was decreased to ~24 hours at RT (Supplementary Fig. 7) which was further decreased to ~30 mins at 40 °C (Supplementary Fig. 8), resulting from the thickness and temperature dependence of the nucleation process. Moreover, the pulse energy used to amorphize the Sb also affects the retention time of a-Sb. While any pulse energy above a threshold can amorphize the Sb, a higher energy (below the damage threshold of the sample) typically gives less nucleation density accounting for a longer retention time for a-Sb (Supplementary Fig. 6 and 7).

**Discussion**

Our experimental results have led to some very interesting observations, namely:

(1) Compared to as-deposited thin film Sb that is very stable (> 6 months at RT), the optically amorphized a-Sb shows a much stronger tendency to recrystallize at RT. If the recrystallization of a-Sb is driven by the growth, it bypasses the nucleation (and speeds up the crystallization) that is required for the crystallization of as-deposited Sb films. On the other hand, if nucleation plays an important role, the optically switched a-Sb contains subcritical nuclei that facilitate the recrystallization[51], leading to a shorter retention time of a-Sb than the as-deposited Sb.

(2) It is known that the reduced glass-transition temperature $T_{rg} = T_g/T_m$ ($T_g$ and $T_m$ are glass transition and melting temperatures respectively) is inversely related to the nucleation rate of PCMs[19,51]. Therefore, we can calculate $T_{rg}$ for Sb as 0.44, with $T_g$ = 400 K (Supplementary Fig. 9) and $T_m$ = 903.5 K[52]. This value is smaller than for $Ge_2Sb_2Te_5$ ($T_{rg}$ = 0.47) and doped Sb ($Ge_{12}Sb_{88}$, $T_{rg}$ = 0.53)[53], qualitatively indicating that Sb has a faster crystallization speed than conventional PCMs.

(3) Thickness-dependent crystallization speed and temperature have been reported in other PCMs[54-56], which is explained by a qualitative model[55] analyzing the energy barrier $E_B$ for crystallization which determines the growth velocity of crystallites. The energy barrier $E_B$ includes the crystalline-amorphous interfacial energy ($E_{ca}$) and the crystalline-interface/surface energy ($E_{ci}$). For a Sb film with thickness $t$, the initial growth of a crystalline cluster of radius $r$ is dominated by $E_{ca}$, given $r < t/2$. Once the size of the cluster surpasses $t$ ($r \geq t/2$), $E_{ci}$, proportional to the crystalline-interface/surface area $S_{ci} = \pi(r^2 - t^2/4)$, will contribute to $E_B$. Therefore, for a given size of the crystalline cluster, thinner Sb has a larger $E_{ci}$, leading to a stronger inhibition to the crystallization. On the other hand, randomly oriented a-Sb atoms activated by thermal energy will move to find a cluster structure with localized minimum energy for initial nucleation with a preference for internal rather than on the surface or interface nucleation. For ultra-thin a-Sb, the ratio of surface or interface atoms to internal atoms is much larger than that in thick or bulk Sb. This makes ultra-thin a-Sb take a longer time to reach initial nucleation and subsequent crystallization. To fully understand the fundamental mechanism of the phase-transition of Sb, especially the optical fast switching, advanced characterization, such as in-situ TEM[57], femtosecond electron[58] and x-ray[59] diffractions and phonon spectroscopy[60], accompanied by theoretical studies[61-63] are necessary, and are beyond the scope of this study.

The potential applications of thin film Sb in silicon photonics are immense but require further investigation. Current integrated photonic memory elements mostly use GST and are based on the extinction ratio contrast between the amorphous and crystalline phases. Compared to GST, Sb has a larger extinction ratio for both a-Sb and c-Sb; however, the contrast is higher indicating that photonic memory using Sb would have a smaller footprint which is crucial for cyclability and interfacing with electronics. Furthermore, our results indicate that Sb can be switched by a fs laser pulse, portending sub-picosecond timescales on integrated devices. In addition, the wide distribution of the retention time of a-Sb with tunable volatility can be employed in photonic neuromorphic computing, to build photonic synapses (non-volatile) and photonic neurons (volatile) by adjusting the thickness of the material[64].

**Conclusions**

In summary, we have explored the optical properties during the solid-state phase transition of a single metal, Sb and find that its optical properties are surprisingly tunable for a range of optical and optoelectronic applications requiring high-speed switching in a thin-film format. Optical constants (*n* and *k*) have a substantial contrast between the amorphous and crystalline Sb when the thickness is less than ~15 nm. The thickness dependent optical properties of Sb indicate that the interfaces of Sb have a significant effect on the phase transition and optical properties. Electrical and optical methods, through CAFM and fs laser respectively, have been used to switch Sb demonstrating high potential for versatile applications in nanophononics and optoelectronics. In addition, the volatility of the optically switched a-Sb can be modulated by the thickness, temperature and the optical pulse energy for switching, indicating a potential material for synaptic and neuron functionalities, with promising applications in photonic neuromorphic computing, high speed holographic and near-eye displays and any other application that requires an actively tunable optical material with metallic properties.

**Acknowledgments:** The authors acknowledge discussions with J. Tan, N. Farmakidis, G. Triggs, R. Taylor, M. Riede, A. Sebastian and A. Ne. **Funding:** This research was supported via the EPSRC grants EP/J018694/1, EP/M015173/1, EP/M015130/1 and EP/R004803/1. Z.C. acknowledges support from Thousand Youth Talents Plan of China. **Author contributions**: All authors contributed substantially. H.B. led the project and with Z.C. conceived and planned the experiments. Z.C., T.M. and S.H. fabricated the samples, performed the AFM, optical and Raman characterizations. Z.C., P.S. and T.M. carried the fs laser experiments under M.B.'s supervision. Z.C., J.K. and T.M. performed the TEM characterizations. Z.C. and T.M collated and analyzed the experimental data. All authors discussed the results and contributed to the manuscript. **Competing interests**: H.B. notes that he serves on the board of directors of Bodle Technologies Ltd. Other authors declare that they have no competing interests. **Data and materials availability**: All data needed to evaluate the conclusion in the paper are present in the paper and/or the Supplementary Materials. Additional data related to this paper may be requested from the authors.

**Methods**

**Film deposition**

Thin films were deposited on silicon wafers (IDB Technologies) from commercial targets (99.99% pure, Testbourne) using RF sputtering (Nordiko sputtering system). Sb films were sputtered at low RF power (30 W) and low pressure (5 mTorr) Ar atmosphere with a deposition rate of 3.33 nm/min. Pt mirror was prepared by sputtering 100 nm Pt on silicon wafers at 50 W, 38 mTorr (8.6 nm/min) with a 5 nm Ta as the adhesion layer. ITO was sputter at 30 W, 5 mTorr, 2.28 nm/min.

**Material characterizations**

**Reflection spectrum** in Fig. 3 was measured using a UV-VIS-NIR spectroscopy (Lambda 1050, PerkinElmer) fitted with a reflectance unit at an angle of incidence of 6 degrees. Local reflection measurements in Fig. 5 was performed with a customized microscopy system, where a white light source was focused on the sample through a 20× objective lens (Thorlabs) with the reflection light collected by a single mode fiber (M15L02-∅105 μm, Thorlabs) and detected by a portable spectrometer (OCEAN-FX-VIS-NIR, Ocean Optics). This customized microscopy system was also employed to determine the crystallization temperature of Sb films on an in-situ heating substrate (Supplementary Fig. 9). **Ellipsometry measurement** was implemented by a spectroscopic ellipsometer (RC2, J.A. Woollam) at three different incident angles. Refractive index and extinction ratio were obtained by fitting measurement results using a built-in software CompleteEASE (J.A. Woollam). **Raman spectrum** was measured by LabRAM ARAM1S (Horiba) using a 532 nm laser with a 50× objective lens, an 1800 grating and a 25% filter. **TEM characterization** was taken by a $LaB_6$ 200 kV transmission electron microscope (JEM-2100, JEOL) at the David Cockayne Centre for Electron Microscopy.

**Electrical and optical switching**

An AFM (MFP-3D, Oxford Instruments Asylum Research) accompanied by a conductive diamond coated tip (DDESP-FM-V2, Bruker) was used to electrically switch the Sb thin film sandwiched between ITO layers. For local switching of Sb, $V_B$ was swept from 0 V to 5 V then back to 0 V while the current $I_S$ passing through Sb was recorded. To switch a large area of Sb, the sample clamped on the piezo stage of the AFM was scanned at 1 kHz with a resolution of 512 points/line. The AFM tip was working in the contact mode with the biased voltage ranging from a minimum (0 V) to maximum (6-8 V) value, corresponding to the grey scale value of the reference image used. For optical switching, a regeneratively amplified Ti: Sapphire laser (Solstice Ace, Spectra Physics) was the switching source, working at the wavelength λ of 790 nm and 1 kHz repetition rate with a pulse duration at the sample of ~200 fs. A single pulse fs laser with the energy $E_p$ = 0.31~1.0 nJ was chosen to amorphize Sb sample. For the recrystallization of the amorphous Sb, 3000 consecutive pulses (at 1 kHz repetition rate) with individual pulse energy of $E_p$ = 0.16 nJ (total energy of 480 nJ per spot) were used, which required much slower translation of the sample (< 0.3 μm/s). To save the total writing time, the laser source was switched to the oscillator that has a much higher repetition rate (80 MHz) but lower pulse energy ($E_p$ = 29 pJ).